\documentclass[twocolumn]{revtex4-2}

\usepackage[utf8]{inputenc}

\usepackage{amsmath}
\usepackage{hyperref}
\usepackage{graphicx}
\usepackage{physics}
\usepackage{bbm}
\usepackage{url}
\usepackage{xcolor}
\hypersetup{
	colorlinks = true,
	linkcolor  = blue,
	citecolor  = blue,
	urlcolor   = blue,
}

\usepackage[normalem]{ulem}

\usepackage{lineno}

\graphicspath{{Figures/}}

\begin{document}
\title{Un-symmetric photon subtraction: a method for generating high photon number states and their relevance to loss estimation at ultimate quantum limit}
\author{N.~Samantaray}
\email{Email: ns17363@bristol.ac.uk}
\author{J. C. F.~Matthews}
\author{J. G.~Rarity}

\affiliation{Quantum Engineering Technology Labs, H. H. Wills Physics Laboratory and
Department of Electrical and Electronic Engineering, University of Bristol, BS8 1FD, UK}

\begin{abstract}
We have studied theoretical un-symmetric multi-photon subtracted twin beam state and demonstrated a method for generating states that resembles to high photon number states with the increase in the number of subtracted photons through Wigner distribution function, which can be reconstructed experimentally by Homodyne measurement. A crucial point is high non-classicality is obtained by photon subtraction when mean photons per mode of twin beam state is low. We have calculated photon statistics from the phase space distribution function and found sub-poissonian behaviour in the same low mean photons regime. Furthermore, we have tested the usefulness of such states for realistic absorption  measurement including detection losses by computing quantum Fisher-Information from measured Wigner function after interaction the sample. We have compared the performance of these states with respect to coherent and demonstrated how the quantum advantage is related to non-classical enhancement. We presented results up to three photon subtraction which show remarkable quantum advantage over both initial thermal and coherent state reaching the ultimate quantum limit in the loss estimation.

\end{abstract}
\maketitle
\section{Introduction}
Photon shot noise limits the precision of measuring a parameter by inverse square root of the number of used photons, also known as shot noise limit in quantum metrology\cite{Dowling:2008}. A major challenge in optical measurement is to search for suitable probe states for overcoming this limit. Squeezed states of light and the so called N00N can possibly allow reaching the fundamental HL in phase estimation \cite{Caves:1981,Aasi:2013,Huver:2008, Benmousa:2003,Louiz:2006}. However, there are two main problems pertaining to their real application, i.e, generation of bright squeezed vacuum \cite{Eberle:13} and high photon N00N \cite{Da:2011,Davidovitch:2011}. The latter is also highly fragile to losses that a loss of single photon mittigatess all the gained advantage. Therefore, demonstration of advantages using such states remains as proof of principle up to now.\par Another most efficient strategy used for sub-shot noise phase and absorption estimation is quantum correlation present in TBS which can be obtained from routinely generated spontaneous parametric down conversion (SPDC) and four wave mixing (FWM) \cite{Berchera:2015,Nigam:2017}. The phase estimation is a unitary process whereas the latter is non-unitary photon loss process leading to ultimate quantum limit \cite{Monras:2007,Losero:2018}. It is paramount to mention SSN advantages using this strategy so far are realised only for low mean number of probe photons per mode ( only mean photons per mode up to 3 can be acieved experimentally), which can be further incresed by photon subtraction \cite{Carranza:12,Chekova:2016}. In addition to entanglement enhancement, photon statistics of individual beams (originally super-poissonian) of TBS not only becomes sub-Poissonian by an action of symmetrical photon subtraction, but it also increases the average number of photons in the resulting state. Nevertheless, the result of symmetrical photon subtraction for absorption measurement does not show much difference with respect to its no subtraction counterpart when the average number of probe photons are balanced \cite{Noi:2020}. Originally, the idea behind symmetrical photon subtraction in TBS is to make the resulting state resembles to a pair coherent state \cite{GSA:1986}, a high photon number entangled Fock state which are yet to be generated experimentally. For loss measurements, Fock states having fixed number of photons approach Uql unconditionally \cite{Adesso:2009}. This optimal limit can be understood from the fact that losses can be estimated by comparing the number of photons both before after interaction with the sample. The knowledge of fixed number of photons of the probe state allows one to estimate very small absorption by the sample, which would remain hidden in the intrinsic photon number fluctuation of Poissonian or thermal distributed sources. The scheme of using high Fock state could allow in principle to reach SSN scaling with high light illumination (high average photon flux) where the advantage due to correlation and entanglement are either not attainable at present or usually lost. Nevertheless, experimentally it is challenging to produce true Fock states with exception of heralded single photon states which are demonstrated for quantum enhanced absorption measuremen both with post-selection of the heralded single photons \cite{Whittaker:2017} and more remarkably, with selection performed by active feed-forward enabled by an optical shutter \cite{Sabines:2017}. Another problem is the availability of high efficient photon number resolving detectors. Although, multi-clicks heralding on idler shows a way for the generation of high Fock state with photon number eqals to number of clicks, the requirement of a very large number of detectors with respect to number of clicks make the scheme  impracticable \cite{Sperling:2014}. \par In the last decades, attentions have been devoted for un-symmetric subtraction from TBS showing improvement in logarithmic negativity a measure of entanglement and negative regions in phase space distribution function forming a quantum optical vortex \cite{Agarwal:2011}. Mathematically subtracting photons from one arm of the TBS is equivalent to adding same number of photons to the other arm before the squeezer. The less clear fact up to now is whether un-symmetric photon subtraction can help in the generation of high photon number Fock states. In this letter, we studied with great importance the un-symmetric multi photon subtracted TBS for answering to following fundamental points: is it possible to enhance non-classical effects in terms of photon statistics and high Fock states generation in the signal state by subtracting photons from idler, and whether or not this probe is useful for getting a better SSN scaling compared to existing resources.
\section{Scheme for un-symmetric subtraction and absorption measurements}
\begin{figure}[htb]
	\centering
	\includegraphics[width=\linewidth]{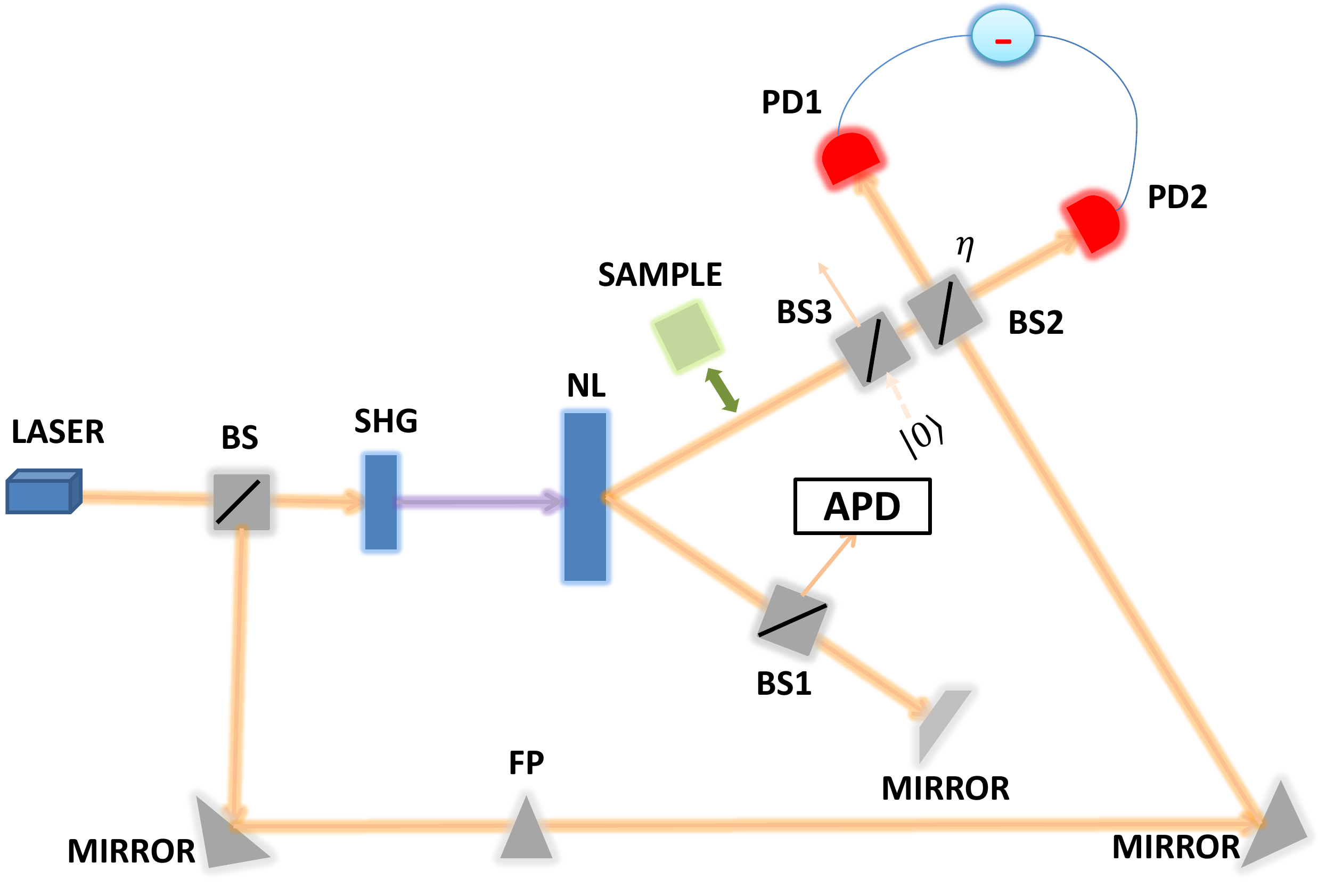}
	\caption{Scheme for un-Symmetric photon subtraction. A laser beam after second harmonic generation (SHG) pumps a non-liner crystal (NL). Photon subtraction is performed by placing a high transmittance beam splitter (BS) on the idler path. A multiplexing channels consisting of more than one single APDs (not shown here) could be useful for differentiating multi-clicks events. Number of clicks on the APDs confirms the same number of photon subtraction and opens the signal path to combine with the a local oscillator(whose phase is varied with the help of a phase plate FP)  at the BS for Homo-dyne measurement and Wigner function is reconstructed for both with and without inserting the sample of absorption coefficient $\gamma$ on the signal path. Detection losses $\eta$ in the Homodyne detectors are modelled by placing a beam splitter $BS3$ of transmittance $\eta$ as shown in the figure \cite{Knyazev:2018}.}
	\label{scheme}
\end{figure}

\begin{figure}[htb]
	\centering
	\includegraphics[width=0.9\linewidth]{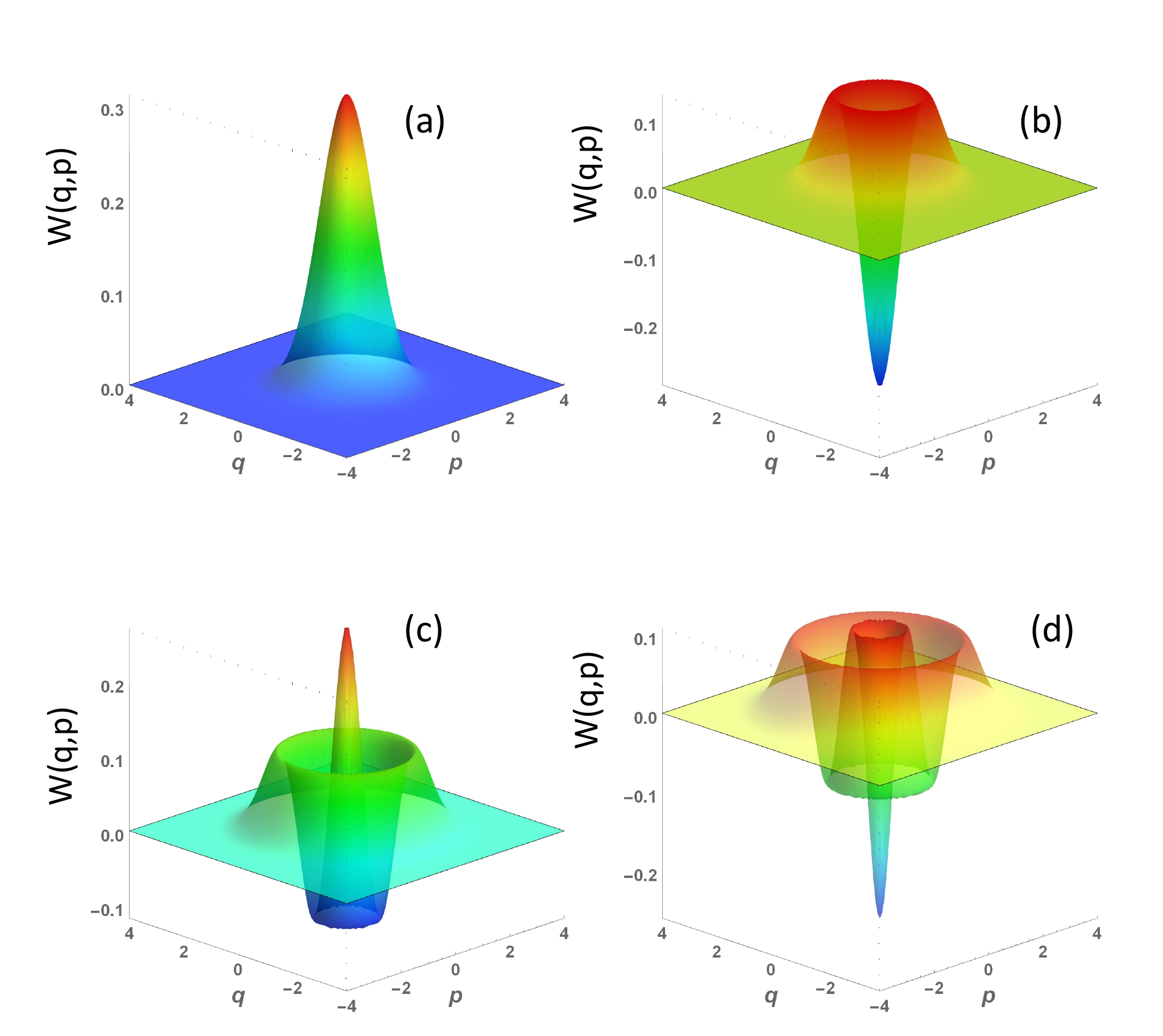}
	\caption{Measured Wigner distribution function at no absorption. The parameter values choosen are $\lambda=0.01$, $\eta=0.98$ and (a) $m=0$, (b) $m=1$, (c) $m=2$ and (d) $m=3$}. 
	\label{zero absorption}
\end{figure}

\begin{figure}[htb]
	\centering
	\includegraphics[width=0.9\linewidth]{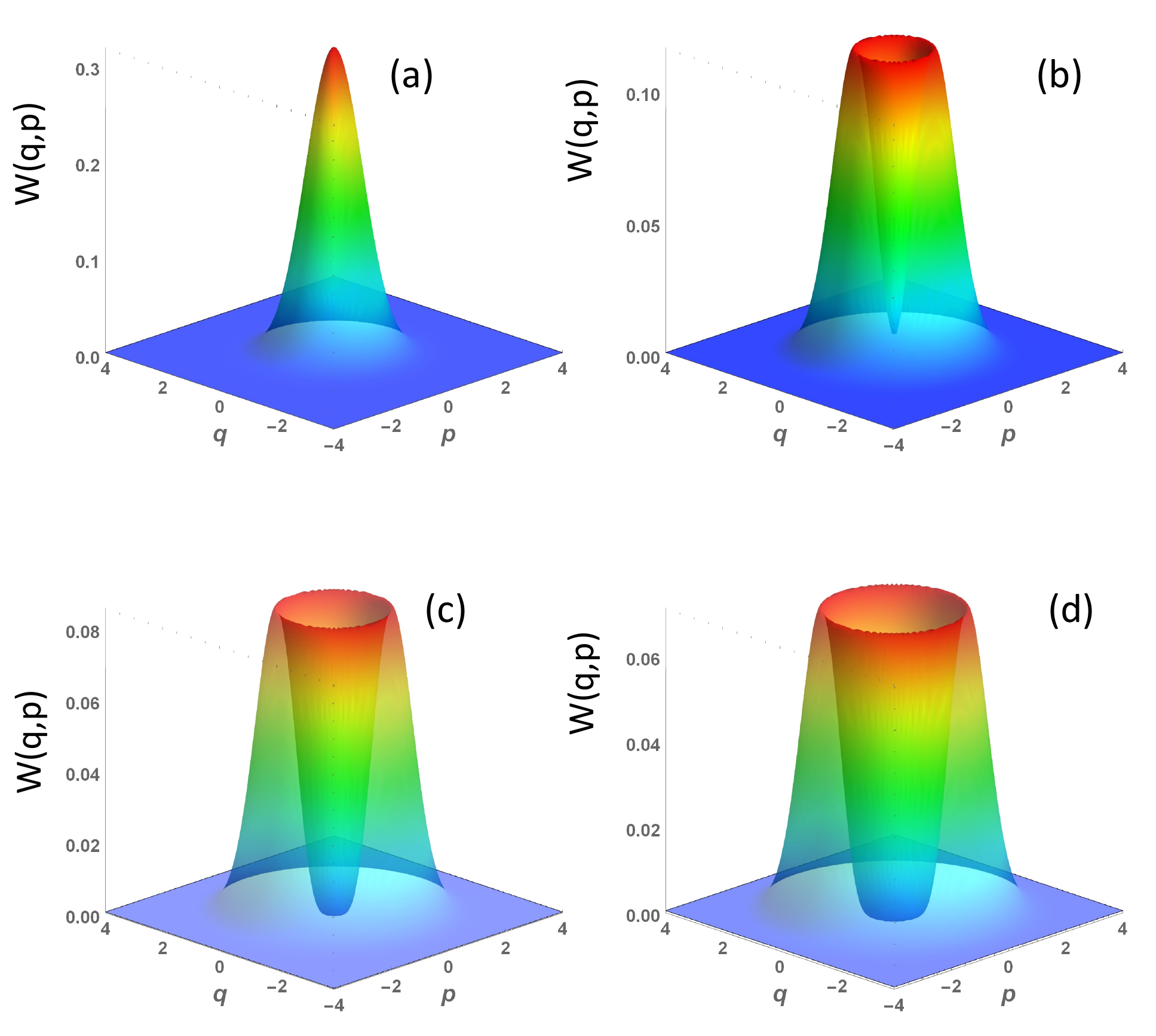}
	\caption{Evolution of Wigner distribution function showing quantum to classical transition in loss process. The parameter values choosen are $\lambda=0.01$, $\eta=0.98$ and (a) $m=0, \gamma=1$, (b) $m=1, \gamma=0.5$, (c) $m=2, \gamma=0.47$ and (d) $m=3, \gamma=0.44$}. 
	\label{arbitrary absorption}
\end{figure}

\begin{figure}[htb]
	\centering
\includegraphics[width=0.9\linewidth]{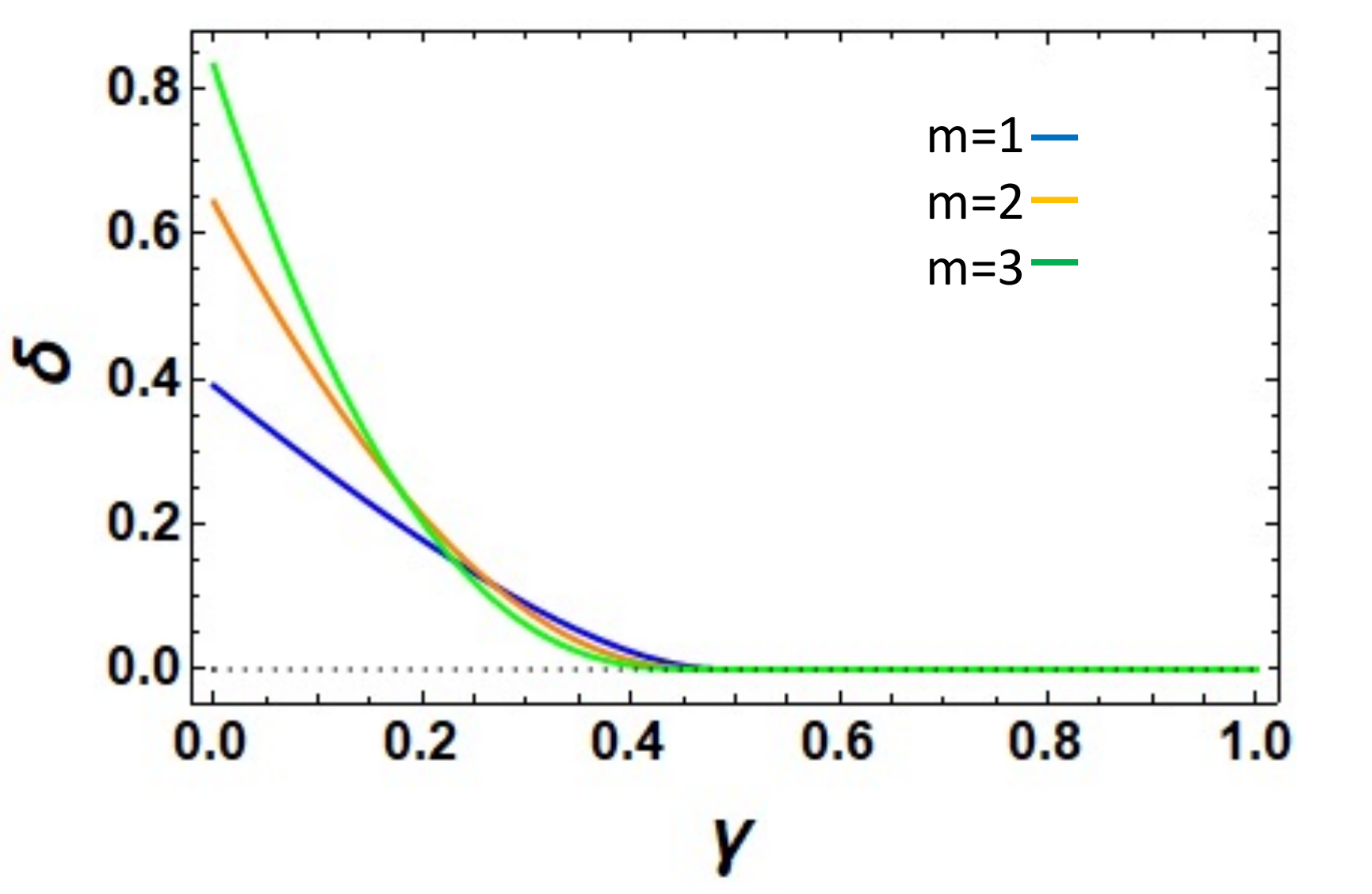}
	\caption{Wigner negative volume versus $\gamma$ for $\lambda=0.01$ and $\eta=0.98$, black dotted line is for coherent state.}. 
	\label{negativevolume}
\end{figure}

To this aim, we proposed an experimentally viable scheme (see Fig. \ref{scheme}) for the generation of such states and their detection from reconstructed phase space function by Homo-dyne measurement varying the phase of a local oscillator.
The whole beam after performing $m$ un-symmetric photon subtraction from idler arm of TBS can be written in the photon number basis as
\begin{eqnarray}\label{Without tracing}
\vert\Psi\rangle^{m}&=&\left(\frac{1}{\sqrt{1+\lambda}}\right)^{m+1}\sum^{\infty}_{n=0}\frac{\sqrt{(m+n)!}}{\sqrt{m!n!}} \\ && \nonumber\times\left(e^{i\chi}\sqrt{\frac{\lambda}{1+\lambda}}\right)^{n}\vert m+n,n\rangle_{s,i},
\end{eqnarray}
where $\lambda=\sinh^{2}r$ is mean photons per mode before subtraction, $r$ being the squeezing parameter and $\chi$ is squeezing angle whose value we have set to zero for simplicity.
Tracing out the idler degrees of freedom from the density matrix of the entire state $\hat{\rho}^{m}_{s,i}=\vert\Psi\rangle^{m}\langle\Psi\vert^{m}$, the signal state becomes
\begin{equation}\label{With tracing}
\hat{\rho}_{s}^{m}=\frac{\left(1-x\right)^{m+1}}{m!}\frac{d^{m}}{dx^{m}}\left(\frac{1}{1-x}\hat{\rho}_{th}\right),
\end{equation}
where $x$ and $\lambda$ are related by $x=\frac{\lambda}{1+\lambda}$ and $\hat{\rho}_{th}$ is a thermal state. Thus, multi-photon subtraction from idler arm resembles to multi-photon added thermal state in the signal arm alone (see supplementary section). It is known that photon addition brings non-classical effects in Gaussian states such as coherent and thermal.In this context, it is intriguing to see the non-classical effects of such states and how their non-classical effects are affected in photon loss channel via Wigner distribution function (WDF). We introduced loss by placing a sample of absorpton coefficient $\gamma$ on the signal path which is equivalent to a beam splitter of transmittance equals to $1-\gamma$ mixed with vacuum state at its free port. Detectors are not ideal and likewise we modelled detection losses by placing another beam splitter of transmittance $\eta$ so that the equivalent transmittance becomes $\tau=\eta(1-\gamma)$. Compact form of WDF in position  (q)-momentum (p) after placing the sample on the signal path is the following (see supplementary section):
\begin{eqnarray}\label{WDFSample}
	W(q,p)&=&\frac{\left(1-x\right)^{m+1}}{\pi m!}\times \\ && \nonumber\frac{d^{m}}{dx^{m}}\left(\frac{1}{1+x(2\tau-1)}\exp\left[\frac{-\left(q^{2}+p^{2}\right)\left(1-x\right)}{1+x(2\tau-1)}\right]\right).
\end{eqnarray}
Wigner function at no absorption shows negative region in phase space and for low values of $\lambda$, the signal resembles to Fock state. More is the number of clicks registered on the APDs, higher is the generated Fock state (see Fig. \ref{zero absorption}). Negative region deteriorates at high $\lambda$ because of dominance from thermal contribution as expected. 
Losses are the main culprit which essentially kills the non-classical effects. The values of $\gamma$ above which non-classicality subsides and the states becomes classical are different for different $m$, and they follow a descending order as the number of subtracted photons $m$ increases (see Fig. \ref{arbitrary absorption}). Another point to note in this classical regime is the phase space area, which is more  for high $m$ compared to its low values in order to normalise the total probability to one.
Quantitatively, the effect of $\gamma$ on non-classicality becomes more clear in Wigner negative volume $\delta=\int \vert{W(q,p)}\vert dpdq$, a positive valued region. For instance at low absorption, the generated three photon subtracted state is nearly twice non-classical compared to $m=1$ (see Fig. \ref{negativevolume}). Classicality for different $m$ correspond to zeros of their negative volume. Thus, these states preserve their quantumness differently to different amount of photon losses, i.e, $m=1$ withstands more absorption loss compared to $m=3$ entering to classical regime in phase space. 
It is possible to obtain photon statistics from WDF through symmetrical ordered product of creation and annihilation operators. Mean number of photons $\langle n\rangle=(1/2)\int (q^{2}+p^{2}) W(q,p)dpdq-(1/2)$ for $m$ photon subtracted state is $(m+1)\lambda+m$. Increase in the definite number of photons are evident for low values of $\lambda$ which is consistent with the results of phase space distribution. Mean square  photon number operator follows from WDF via $\langle n^{2}\rangle=(1/4)\int (q^{2}+p^{2})^{2} W(q,p)dpdq-(1/2)\int (q^{2}+p^{2}) W(q,p)dpdq$ from which variance in $\hat{n}$, Var($\hat{n})=(\langle n\rangle)^{2}-\langle \hat{n}^{2}\rangle$ and photon statistics in terms of Fano factor $F=\frac{Var(\hat{n})}{\langle n\rangle}$ can be easily obtained. Intuitively, Fano factor should not show any difference in the generated states for different $m$ at very low $\lambda$ unlike the negative volume. As $\lambda$ increases, thermal contribution adds up changing the Fock state statistics ($F=0$) progressively to super-poissonian ($F>1$), Poissonian statistics corresponds to $F=1$. However, sub-poissonian statistics ($F<1$) is retained up to certain $\lambda$ and the ranges varies for different $m$ (see figure 5).
\begin{figure}[htb]
	\centering
	\includegraphics[width=0.8\linewidth]{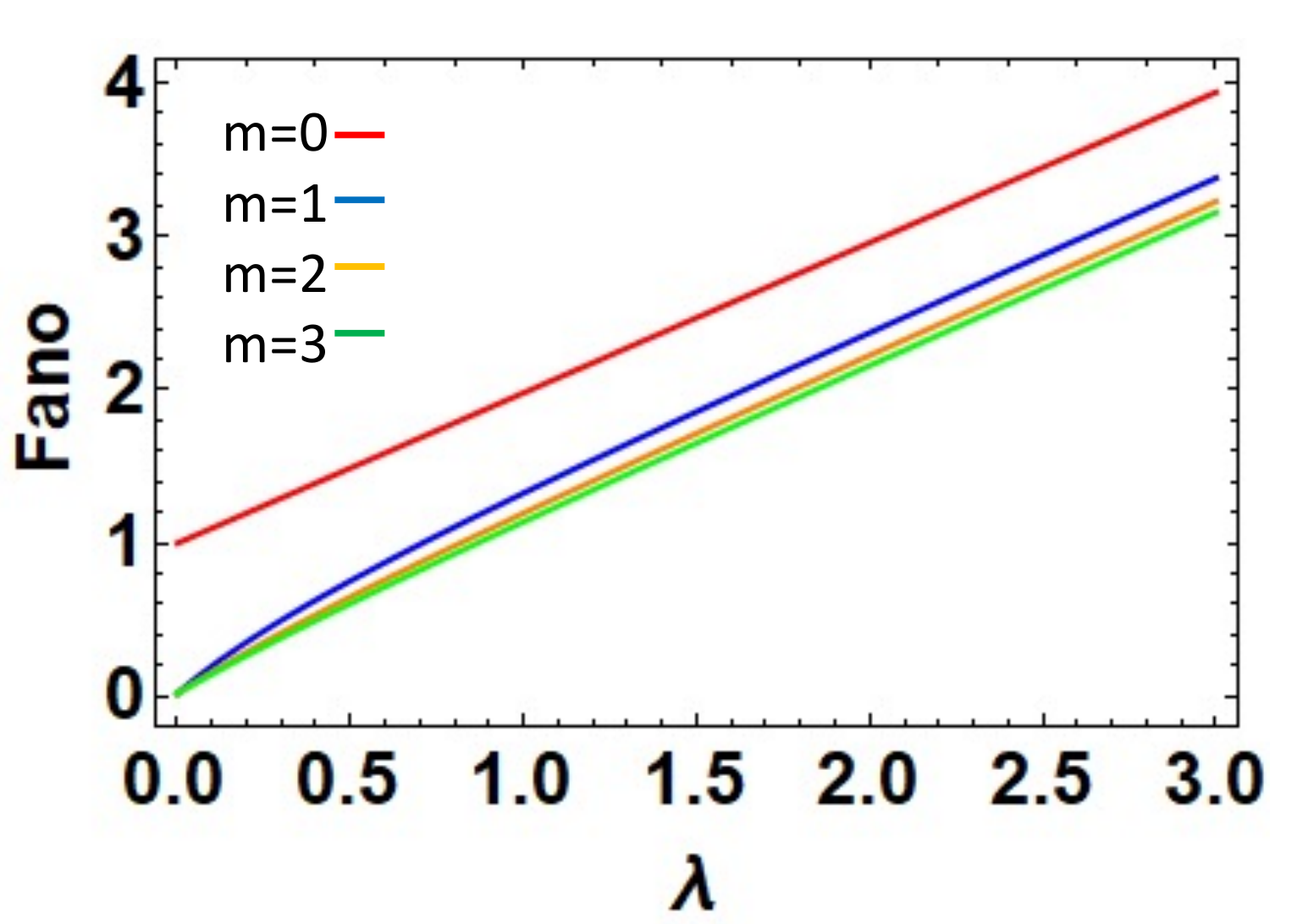}
	\caption{Measured Fano Factor versus $\lambda$ for $\eta=0.98$.}. 
	\label{Measured Fano factor}
\end{figure}
Increase in the mean number of photons, enhancement in non-classicality and sub-poissonian statistics due to unsymmetrical photon subtraction are the subject of this work and can be useful for absorption measurement. Possibility of using such states for loss estimation is explored in terms of quantum fisher information (QFI). For pure state probe, computation of QFI for a parameter to be estimated involves measurement of variance associated with its generator. However finding QFI in decoherence free space is a hard task in quantum metrology. We have brought a method for getting QFI in loss estimation from experimental measured WDF after passing through the sample regardless of the process. This approach is different from the reported pure state case where QFI is computed from theoretically obtained Wigner functions both before and after the unitary process \cite{Braun:2014}. The principal points about our approch are: WDF carries information about both position and momentum degrees of freedom which are implicit function of the parameters to be estimated. Integrating over one degrees of freedom results a probability distribution for the other carrying information about the parameter and vice versa, and from this probability distribution, an experimental measure of QFI can be obtained (see supplementary section). Following this approach, QFI for absorption measurement is obtained (see Fig. \ref{QFIgamma}) for different $m$ (solid lines). \par Performances of Fock states up to three photons are shown by the corresponding dotted lines (imrpovement in uncertainty gets smaller for each increase in $m$) which gives clear evidences about the following: as long as detection efficiency is high, the generated states by $m$ photon subtraction (except for $m=0$ case) and the Fock states almost give similar advantage and carry more information about the object with respect to coherent state at low $\lambda$. The advantage is highest at low $\gamma$. At high absorption, QFI of the generated signal states and Fock states tend to reach asymptotically the SNL (see supplementary section for coherent state performance for loss estimation) as expected. Aside magnitude, quantum advantage of different $m$ follow similar behaviour like negative volume against $\gamma$ before going to the classical regime. The only difference is that their magnitudes in the classical regime (both subtracted states and Fock states look one and the same) are still higher than coherent state in contrast to negative volume as QFI is related to mean number of photons and statistics. QFI for $m=0$, though, gives an impression of getting zero information in absorption measurement, which is not true in general as we shall see in some other regime of interest and for other values of mean energy $\lambda$ (as magnitude tends to zero for nearly zero mean energy).
\begin{figure}[htb]
	\centering
	\includegraphics[width=0.9\linewidth]{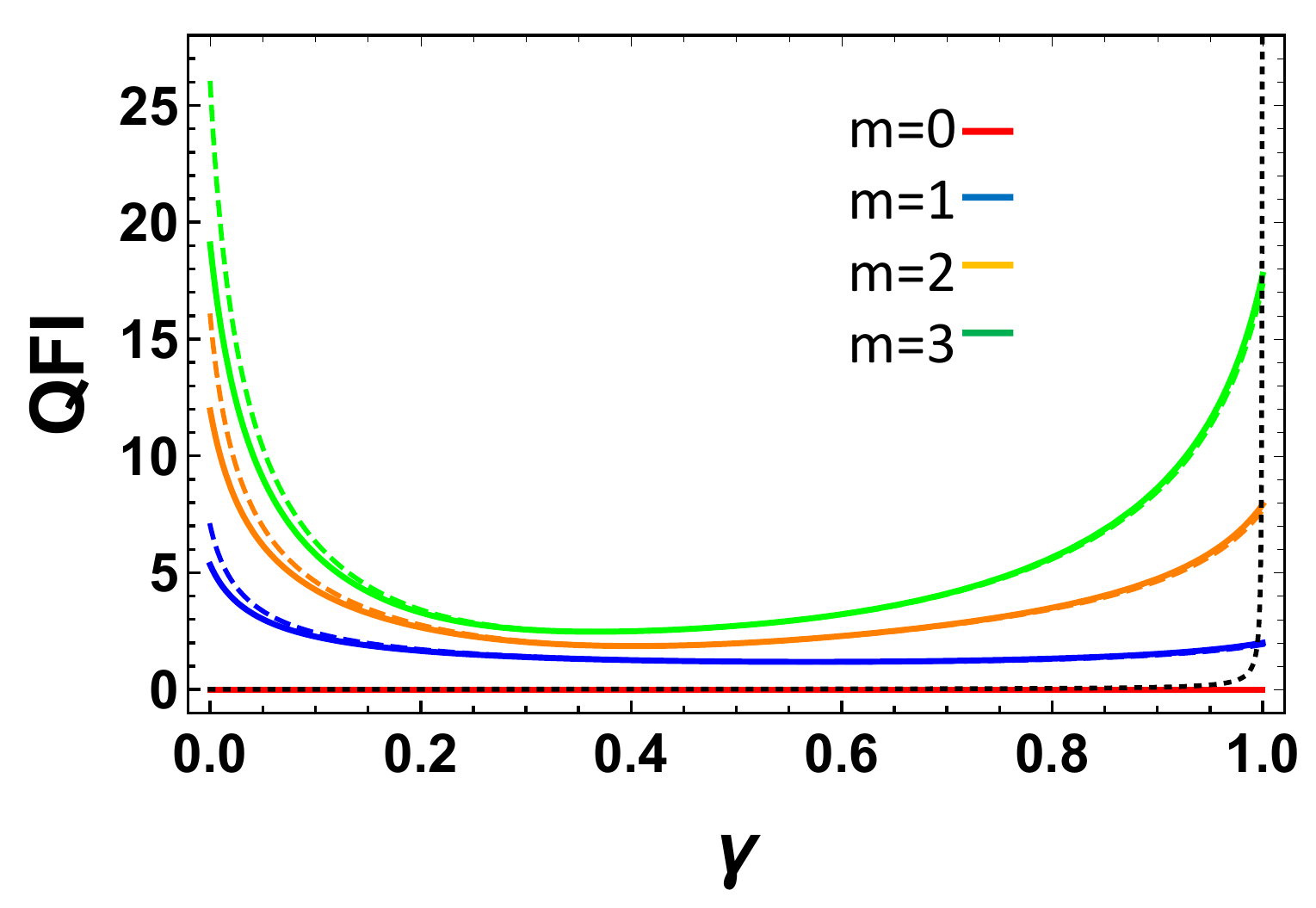}
	\caption{QFI versus $\gamma$ for $\eta=0.98$ and $\lambda=0.01$, dashed lines correspond to performance of Fock states and black dotted line is coherent state with mean photon $\mu=\lambda$}. 
	\label{QFIgamma}
\end{figure}
\par
\begin{figure}[htb]
	\centering
	\includegraphics[width=0.9\linewidth]{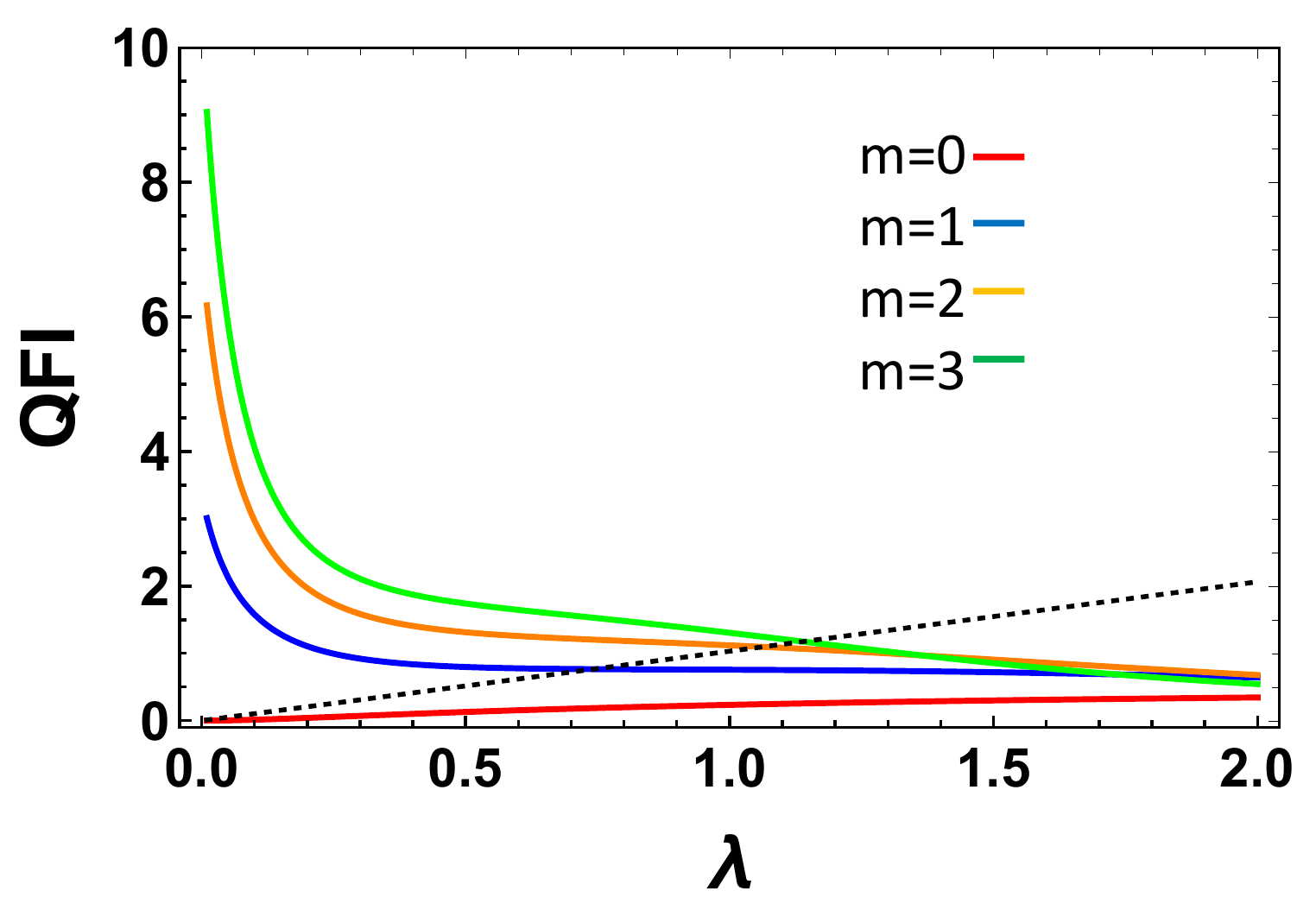}
	\caption{QFI versus $\lambda$ for $\eta=0.98$ and $\gamma=0.01$, dotted line is coherent state with mean photon $\mu=\lambda$}. 
	\label{QFIlambda}
\end{figure}

 Variation of QFI against $\lambda$ has been shown in Fig. \ref{QFIlambda} for low  $\gamma$. Fock states performance (not shown) does not change regardless of  $\lambda$ as they are independent of it. Only at low  $\lambda$, the subtracted states perform equally well with respect to Fock states as expected. As  $\lambda$ increases, thermal contribution in $m$ photon subtracted states increase resulting in a dominating performance of coherent state after certain  $\lambda$ whose threshold value increases with $m$.
Detection and absorption losses do not commute in general. We have checked that in the same  high $\lambda$ regime, $m$ photon subtracted states outperforms both coherent state and Fock states at considerably high values of detection losses.
\section{Conclusions}
We have studied a situation where photons are subtracted from one arm of TBS. We have demonstrated photon subtraction in one arm not only leads to sub-poissonian photon statistics in the other but it also helps in the generation of Fock states at low mean energies of TBS. Interestingly, the resulting photon subtracted state after tracing out the idler turns to a multi-photon added thermal state. We developed a general but simple noble scheme demonstrating the usefulness of such states for loss estimation in terms of quantum fisher information from their experimental measured WDF after passing through the sample. Furthermore, our scheme shows a way to transit from quantum to classical regime both in terms of WDF and negative volume when losses are increased beyond certain threshold. We found twice non-classical enhancement for three photon subtraction compared to single photon. Moreover we have demonstrated the improvent at low loss estimation is linked to non-classical enhancement of  such states. Our strategy is useful for SSN measurement with a capability of generating high Fock states which was not possible for TBS due to it low mean photons per mode. Photon subtraction advantage can be attributed to enhancement in non-classicality, increase in the mean photons and improvement in the photon statistics. At low mean photons per mode, though improvement in the measured uncertainty due to higher order photon subtraction decreases for each increase in the number of subtracted photon, they tend to reach the ultimate quantum limit. 

\section{Supplementary section}
\subsection{Photon subtracted TBS:}
TBS which is theoretically obtained by applying two mode squeezing operator, i.e,
\begin{equation}
\vert\Psi\rangle=\hat{S}_{s,i}(\lambda e^{i \chi})\vert 0,0\rangle	
\end{equation}
Since photon subtraction is a non-unitary operation, the resulting state after $m$ photon subtraction is
 \begin{eqnarray}
 	\vert\Psi\rangle^{m} &=& N^{m}\hat{a}_{i}^{m}\hat{S}_{i,s}(\lambda)\vert 0,0\rangle  \\  \nonumber &=& N^{m}\hat{S}_{i,s}(\lambda)\hat{S}^{\dagger}_{i,s}(\lambda)\hat{a}_{i}^{m}\hat{S}_{i,s}(\lambda)\vert 0,0\rangle,
 \end{eqnarray}
where $N^{m}$ is a normalization constant. In the second step, we have used the identity $\hat{S}_{i,s}(\lambda)\hat{S}^{\dagger}_{i,s}(\lambda)=\hat{I}$. Using the two mode squeezing operator transformation  ${S}^{\dagger}_{i,s}(\lambda)\hat{a}_{i}\hat{S}_{i,s}(\lambda)=\hat{a}_{i}\sqrt{1+\lambda}+\hat{a}^{\dagger}_{s}e^{i\chi}\sqrt{\lambda}$, and as the two modes are independent, the normalization constant easily follow
\begin{equation}
N^{m}=\frac{1}{\sqrt{m!}(\sqrt{\lambda})^{m}}.	
\end{equation}
The resulting state after photon subtraction follows
\begin{equation}
\vert\Psi\rangle^{m}=\hat{S}_{i,s}(\lambda)\vert 0,m\rangle.	
\end{equation}
This shows subtracting photons from one mode is equivalent to seeding a corresponding Fock state to other mode before the squeezing operation. Writing $\hat{S}_{i,s}(\lambda)$ in the normal order form and using the fomulae of creation and annihillation operator on Fock state, 
the unsymmetric photon subtracted state is 
\begin{eqnarray}
	\vert\Psi\rangle^{m}&=&\left(\frac{1}{\sqrt{1+\lambda}}\right)^{m+1}\sum^{\infty}_{n=0}\frac{\sqrt{(m+n)!}}{\sqrt{m!n!}} \\ && \nonumber\times\left(e^{i\chi}\sqrt{\frac{\lambda}{1+\lambda}}\right)^{n}\vert m+n,n\rangle_{s,i},
\end{eqnarray}
Tracing out the idler mode from the entire two mode state and letting $x=\frac{\lambda}{1+\lambda}$, density matrix of signal becomes
\begin{eqnarray}\label{photon added thermal}
\hat{\rho}^{m}_{s} &=&Tr_{i}\left[\hat{\rho}^{m}_{i,s}\right] \\  \nonumber &=& \left(1-x\right)^{m+1}\sum^{\infty}_{p=0}\frac{(p+m)!}{m! p!}\left(x\right)^{p}\vert m+p\rangle\langle m+p\vert.
\end{eqnarray}
The above summation can be further simplified by letting $s=p+m$ and changing the summation index from $p$ to $s$, the state becomes
\begin{eqnarray}\label{photon added thermal1}
\hat{\rho}^{m}_{s} &=&\frac{\left(1-x\right)^{m+1}}{m!}\sum^{\infty}_{s=0}s(s-1)...(s-m+1)\\ && \nonumber \times\left(x\right)^{s-m}\vert s\rangle\langle s\vert \\  \nonumber &=& \frac{\left(1-x\right)^{m+1}}{m!}\frac{d^{m}}{dx^{m}}\left[\frac{\hat{\rho}_{th}}{1-x}\right],
\end{eqnarray}
where $\hat{\rho}_{th}$ is a thermal state 
\begin{equation}\label{thermal}
\hat{\rho}_{th}=(1-x)\sum^{\infty}_{j=0}x^{j} \vert j\rangle\langle j\vert,
\end{equation}
with thermal coefficient $x$ as $s$  is dummy index in the summation.
Eq. \ref{photon added thermal} looks like a photon added thermal state.

\subsection{Beam splitter transformation of $\hat{\rho}^{m}_{s} :$}
\begin{figure}[htb]
	\centering
	\includegraphics[width=8cm,height=5.6cm]{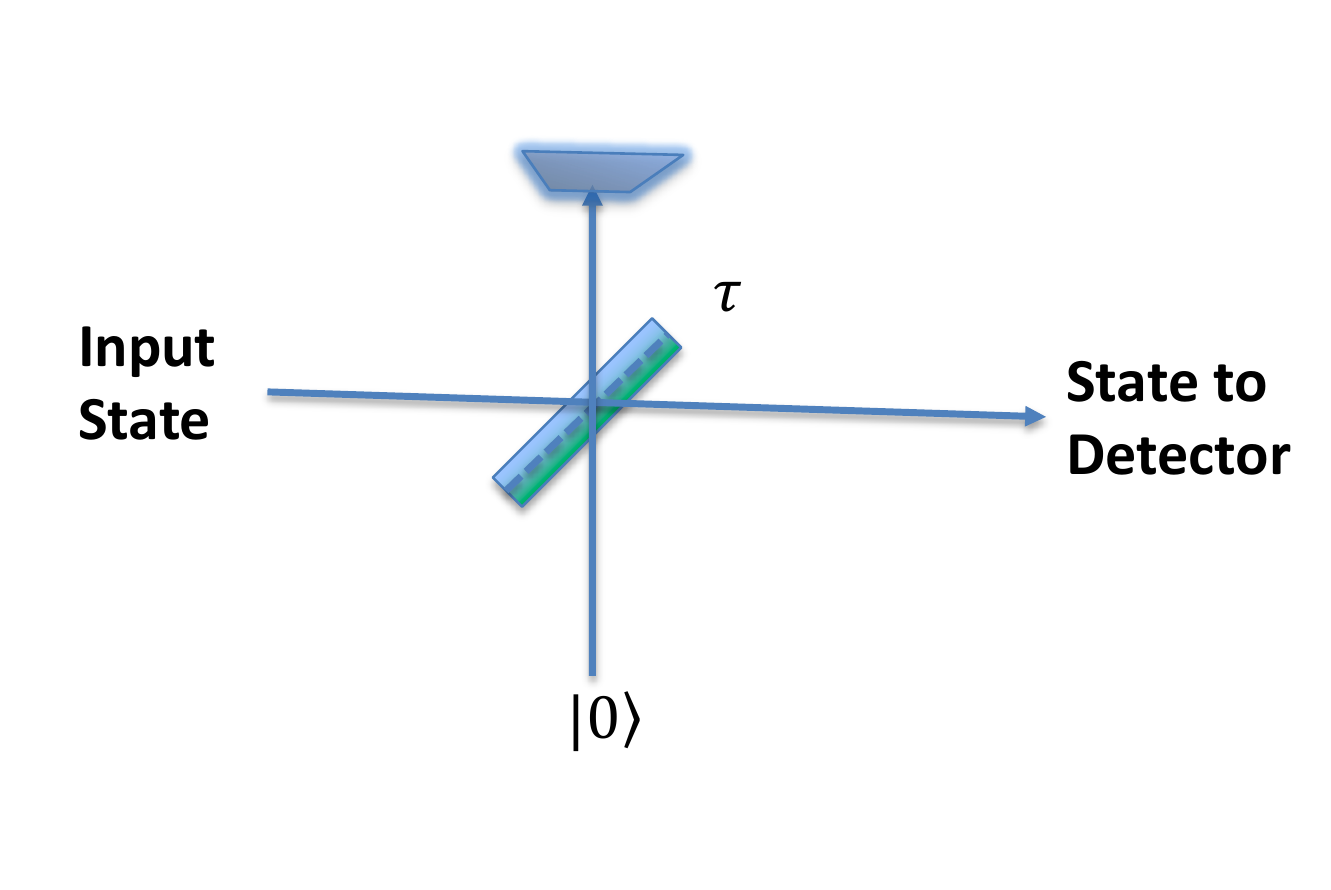}
	\caption{Schematic of beam splitter transformation where an input quantum state is mixed with vacuum and the transmitted part is sent for detecton}. 
	\label{beam splitter}
\end{figure}

A general picture showing trandformation of arbitrary quantum state is depicted in the Fig. \ref{beam splitter}.
For Fock state $\vert j\rangle$ input, the total state at the output ports can be written as
\begin{equation}
\vert \Phi\rangle = \sum^{j}_{k=0}\sqrt{\frac{j!}{k!(j-k!)}}(\sqrt{\tau})^{j-k}(\sqrt{\tilde{r}})^{k}\vert j-k,k\rangle_{t,r},
\end{equation}
where $t,r$ are the transmitted and reflected modes respectively. $\tau$ and $\tilde{r}$ are beam splitter transmittance and reflectance so that $\tau+\tilde{r}=1$.

Tracing out the reflection mode, the state transmitted is
\begin{eqnarray}\label{tracedoutbs}
	\hat{\tilde{\rho}}_{t} &=&Tr_{r}\left[\vert \Phi\rangle\langle \Phi\vert\right] \\  \nonumber &=& \sum^{j}_{k=0}\frac{j!}{k!(j-k)!}\tau^{j-k}(\tilde{r})^{k}\vert j-k\rangle\langle j-k\vert.
\end{eqnarray}
Summation index can be changed from $k$ to $l$ by letting $j-k=l$. The above expression can be rewritten as
\begin{equation}
	\hat{\tilde{\rho}}_{t}= \sum^{j}_{l=0}\frac{j!}{l!(j-l)!}\tau^{l}(\tilde{r})^{j-l}\vert l\rangle\langle l\vert.
\end{equation}
Referring to eq.\ref{thermal}, the state transmitted for thermal state as input is
\begin{equation}\label{tracedoutfockbs}
	\hat{\tilde{\rho}}^{(out)}_{t}= (1-x)\sum^{\infty}_{j=0}x^{j}\sum^{j}_{l=0}\frac{j!}{l!(j-l)!}\tau^{l}(\tilde{r})^{j-l}\vert l\rangle\langle l\vert.
\end{equation}
We are interested in finding the phase space distribution of the transmitted state given the input state in eq.\ref{photon added thermal1}.
Wigner distribution function (WDF) of Fock state with density operator $\vert l\rangle \langle l\vert$ is
\begin{equation}
W_{\vert l\rangle\langle l\vert}(q,p)=\frac{(-1)^{l}}{\pi}\exp\left(-q^{2}-p^{2}\right)L_{l}\left(2q^{2}+2p^{2}\right),
\end{equation}
where $L_{l}$ is $l^{th}$ order Laguerre polynomial, and $q$ and $p$ are position and momentum variables respectively.
Letting $\alpha=q^{2}+p^{2}$, WDF of the state given in eq. \ref{tracedoutfockbs} follows easily
\begin{eqnarray}\label{rhoout}
	W_{\hat{\tilde{\rho}}^{(out)}_{t}}(q,p)&=&\frac{(1-x)e^{-\alpha}}{\pi}\sum_{j=0}^{\infty}(x\tilde{r})^{j} \\ && \nonumber\times\left[\sum_{l=0}^{j}\frac{(-1)^{l}j!}{l!(j-l)!}\left(\frac{\tau}{\tilde{r}}\right)^{l}L_{l}\left(2\alpha\right)\right]_{1}
\end{eqnarray}
In order to find a close form expression of the above equation, we use series expansion of Laguerre polynomial, the bracketed part can now be written as
\begin{equation}
[]_{1}=\sum_{l=0}^{j}\frac{(-1)^{l}j!}{l!(j-l)!}\left(\frac{\tau}{\tilde{r}}\right)^{l}\sum_{p=0}^{l}\frac{(-1)^{p}(2\alpha)^{p}l!}{(l-p)!(p!)^{2}}
\end{equation}
Reversing the order of two summation and after doing further simplification, the bracketed part is 
\begin{equation}
	[]_{1}=(-1)^{j}\left(\frac{\tau-\tilde{r}}{\tilde{r}}\right)^{j}L_{j}\left(\frac{2\alpha\tau}{\tau-\tilde{r}}\right)
\end{equation}
Substituting the bracketed part in eq.\ref{rhoout} and further using the summation identity of Laguerre polynomial, i.e $\sum_{j=0}^{\infty}L_{l}(z)w^{j}=\frac{1}{1-w}e^{\frac{wz}{w-1}}$ and setting $\tilde{r}=1-\tau$, we get a close form expression of WDF as follows
\begin{eqnarray}\label{WDFSample1}
W_{\hat{\tilde{\rho}}^{(out)}_{t}}(q,p)&=&\left(\frac{1-x}{1+x(2\tau-1)}\right) \\ && \nonumber\times\exp\left[\frac{-\left(q^{2}+p^{2}\right)\left(1-x\right)}{1+x(2\tau-1)}\right].
\end{eqnarray}
Now for the state in eq. \ref{photon added thermal1}, WDF becomes
\begin{eqnarray}\label{WDFSample1}
	W(q,p)&=&\frac{\left(1-x\right)^{m+1}}{\pi m!}\times \\ && \nonumber\frac{d^{m}}{dx^{m}}\left(\frac{1}{1+x(2\tau-1)}\exp\left[\frac{-\left(q^{2}+p^{2}\right)\left(1-x\right)}{1+x(2\tau-1)}\right]\right).
\end{eqnarray}

\subsection{Quantum Fisher information from measured WDF:}
\subsubsection{Quantum Fisher information for coherent state:}
Following the method of previous section, Wigner function for coherent state $\vert\beta\rangle$ with mean photons $\mu=|\beta|^{2}$ passing through the sample of absorption coefficient $\gamma=\frac{\tau+\eta}{\eta}$ becomes
\begin{equation}
W_{\vert\beta\rangle\langle\beta\vert}(q,p)=\left(\frac{2}{\pi}\right)\exp\left[-2\left(q^{2}+p^{2}-2q\sqrt{\mu\tau}+\mu\tau\right)\right]
\end{equation}
Conditional probability distribution of q given a parameter $\gamma$ can be obtained from WDF by integrating over its momentum variable $p$
\begin{eqnarray}
	Pr\left[q(\gamma)\right]&=&\int_{-\infty}^{\infty}W_{\vert\beta\rangle\langle\beta\vert}(q,p)\partial p\\ &&\nonumber =\sqrt{\frac{2}{\pi}}\exp\left[-2q^{2}+4q\sqrt{\mu\tau}-2\mu\tau\right]
\end{eqnarray}
QFI can be calculated for a given probability distribution $Pr\left[q(\gamma)\right]$ as
\begin{eqnarray}
	QFI &=&\int_{-\infty}^{\infty}\frac{1}{Pr\left[q(\gamma)\right]}\left(\frac{\partial Pr\left[q(\gamma)\right]}{\partial\gamma}\right)^{2}\partial\gamma\\ &&\nonumber =\frac{\eta\mu}{1-\gamma}
\end{eqnarray}
Uncertainty in measuring the parameter $\gamma$ for many iterative measurement becomes
\begin{equation}
	\Delta\gamma\approx\frac{1}{\sqrt{QFI}}=\sqrt{\frac{1-\gamma}{\eta\mu}},
\end{equation}
which is the SNL in absorption measurement. Therefore SSN limit can reached for for probe state when its QFI is higher than that of coherent state.
\subsubsection{Quantum Fisher information for Fock state:}
Wigner function for Fock state $\vert j\rangle$ state passing through the sample is
\begin{eqnarray}
W_{\vert j\rangle\langle j\vert}(q,p)&=&\frac{\left(1-2\tau\right)^{j}\exp\left[-\left(q^{2}+p^{2}\right)\right]}{\pi} \\ && \nonumber \times L_{j}\left(\frac{2\tau\left(q^{2}+p^{2}\right)}{2\tau-1}\right)
\end{eqnarray}
Following same produre mentioned for coherent state, we numerically computed QFI. We have noticed, $\Delta\gamma$ using Fock states reach UQL 
\begin{equation}
\Delta\gamma\approx\frac{1}{\sqrt{QFI}}\approx\sqrt{\frac{\gamma(1-\gamma)}{\eta j}}
\end{equation}
for high  $\vert j\rangle$ in this scheme.
\acknowledgements
The authors would like to thank etc. etc.
(Quantic,)
\bibliography{library.bib}
\end{document}